\documentclass[12pt]{iopart}
\usepackage{rotating}
\usepackage{color}
\begin{document}
\newcommand {\braket}[2]{ \langle #1 | #2  \rangle }
\newcommand {\braHket}[3]{ \langle #1 | #2 | #3 \rangle }
\newcommand {\ket}[1]{| #1 \rangle}
\newcommand {\bra}[1]{\langle  #1 | }
\newcommand \diff {{\rm{d}}}
\newcommand \eye {{\rm{i}}}
\newcommand \sinc {{\rm{sinc\ }}}

\title{Linear-scaling electronic structure theory: Electronic temperature in the Kernel Polynomial Method}
\author{Eunan J. McEniry$^1$, Ralf Drautz$^2$}
\address{$^1$ Department of Computational Materials Design, Max-Planck-Institut f\"ur Eisenforschung GmbH, D\"usseldorf, Germany}
\address{$^2$ Interdiscplinary Centre for Advanced Materials Simulation, Ruhr-Universit\"at Bochum, Germany}
\ead{e.mceniry@mpie.de}
\date{\today}

\begin{abstract}
Linear-scaling electronic structure methods based on the calculation of moments of the underlying electronic Hamiltonian offer a computationally efficient and numerically robust scheme to drive large-scale atomistic simulations, in which the quantum-mechanical nature of the electrons is explicitly taken into account. We compare the kernel polynomial method to the Fermi operator expansion method and establish a formal connection between the two approaches. We show that the convolution of the kernel polynomial method may be understood as an effective electron temperature. The results of a number of possible kernels are formally examined, and then applied to a representative tight-binding model.
\end{abstract}


\maketitle
\section{Linear-scaling electronic structure approaches via the calculation of moments}

The Fermi Operator Expansion (FOE) and the Kernel Polynomial Method (KPM) achieve linear scaling electronic structure computations by an iterative evaluation of a local Hamiltonian. The two methods appear to follow different strategies for the evaluation of forces and energies. The FOE expands the occupation number of the eigenstates represented by a temperature dependent smearing function. If the smearing function, which corresponds to the Heaviside step function at zero temperature, is represented by a smooth expansion, then the resulting density matrix is short ranged and thus local. The smoothing of the step function is expected to be efficient in particular for systems with a band gap at the Fermi level, thus the FOE is often portrayed as being applicable in particular for insulators and semi conductors. In the Kernel Polynomial Method (KPM) the density of states is convoluted with a kernel that leads to a smoothed representation of the density of states. The kernel corresponds to a broadened representation of a Dirac delta function, and a broader kernel implies a more local evaluation of the energy. The KPM obtains energies and forces from the density of states and in this way avoids the explicit computation of the density matrix. Therefore it is often seen as a method that is suitable for simulations in metals with their long-ranged density matrix. The KPM may be related to recursion based methods that also focus on the density of states.\cite{Weisse06,Seiser13}

We will first compare and contrast the FOE and the KPM and then show how the two methods are closely related. From this straightforward analysis the smoothing of the density of states in the KPM may be understood as a temperature broadening, which will then lead us to suggest a new kernel for the KPM. The new kernel performs practically as well as the standard kernel.

In {\it ab-initio} or tight-binding electronic structure calculations, the electronic temperature $T$ is typically introduced by applying a smearing function at the Fermi level. The band energy and electron count are then obtained as
\begin{eqnarray}
N & = &  \int f (\varepsilon, \mu) n (\varepsilon) \diff \varepsilon , \label{eq:nelec}\\
U_{band} & = & \int \varepsilon f (\varepsilon, \mu) n (\varepsilon) d \varepsilon , \label{eq:onsite}
\end{eqnarray}
where $\mu$ is the electron chemical potential, $ f (\varepsilon, \mu)$ the temperature dependent smearing function and $n (\varepsilon)$ the density of states. At $T=0$K the smearing is zero and $f (\varepsilon, \mu)$ corresponds to the Heaviside step function $ \Theta (\varepsilon, \varepsilon_F)$ which is one below the Fermi energy and zero above. In the Fermi operator expansion (FOE) method\cite{Goedecker95,Goedecker99} the density matrix is locally approximated by writing it as
\begin{equation}
\rho_{ij}  = \int f (\varepsilon, \mu) n_{ij} (\varepsilon) d \varepsilon \,, \label{eq:densitymatrix}
\end{equation}
where $n_{ij} (\varepsilon)$ is the spectrally resolved density matrix.
The diagonal elements $n_i (\varepsilon) = n_{ii} (\varepsilon)$ of the spectrally resolved density matrix correspond to the local density of states associated with orbital $\ket{i}$ , such that the number of electrons in this orbital is obtained as,
\begin{equation}
N_i  = \int f (\varepsilon, \mu) n_i (\varepsilon) d \varepsilon \,.
\end{equation}
The moments of the spectrally resolved density matrix and the local density of states may be related to expectation values that can be computed from Hamiltonian matrix elements alone,
\begin{eqnarray}
\xi_{ij}^{(n)}  &= \int \varepsilon^n  \, n_{ij} (\varepsilon) d \varepsilon = \braHket {i} {\hat {H}^n} {j} \,, \label{eq:int}  \\
\mu_i^{(n)}  &= \int \varepsilon^n  \, n_i (\varepsilon) d \varepsilon = \braHket {i} {\hat {H}^n} {i} \,.  \label{eq:mom}
\end{eqnarray}
Therefore, by writing a polynomial expansion of the smearing function
\begin{equation}
f (\varepsilon, \mu) = \sum_k c_k \varepsilon^k \,, \label{eq:poly}
\end{equation}
an expansion of the density matrix may be obtained,
\begin{equation}
\rho_{ij}  = \sum_k c_k \xi_{ij}^{(k)} \,.
\end{equation}
This is the basis for the Fermi operator expansion, where in practice Chebyshev polynomials are used instead of a direct polynomial expansion\cite{Goedecker95}. The band energy is then obtained in intersite representation,
\begin{equation}
U_{band} =  \sum_{ij}  \rho_{ij}  H_{ji}\,,
\end{equation}
which for $T=0$K is equivalent to the onsite representation Eq. (\ref{eq:onsite}).
Clearly the band energy according to Eq.(\ref{eq:onsite}) may also be directly expanded using Eq.(\ref{eq:poly}),
\begin{equation}
U_{band} =  \sum_i  \sum_k c_k \mu_i^{(k+1)}\,,
\end{equation}
which follows directly from Eq.(\ref{eq:mom}). 

In the kernel polynomial method \cite{Silver94, Silver96, Weisse06} one takes a different approach to the calculation of the number of electrons and the band energy. In a first step an approximate density of states $\tilde {n}_i (\varepsilon)$ is obtained from
\begin{equation}
\tilde {n}_i (\varepsilon) =  \int K (\varepsilon, \varepsilon') n_i (\varepsilon') d \varepsilon \,,\label{eq:kpm_definition}
\end{equation}
where $ n_i (\varepsilon)$ is the local density of states and $K (\varepsilon, \varepsilon')$ the kernel.
The kernel is expanded in Chebyshev polynomials, making use of the moments theorem Eqs.(\ref{eq:int},\ref{eq:mom}). A strictly positive representation of $K (\varepsilon, \varepsilon')$ guarantees that the resulting density of states $\tilde {n}_i (\varepsilon)$ is also strictly positive.
The number of electrons and the band energy is then obtained from
\begin{eqnarray}
N_i &=  \int^{\varepsilon_F} \tilde {n}_i (\varepsilon) d \varepsilon =  \int \Theta  (\varepsilon, \varepsilon_F) \tilde {n}_i (\varepsilon) d \varepsilon \,, \\
U_{band,i} &=  \int^{\varepsilon_F} \varepsilon \, \tilde {n}_i (\varepsilon) d \varepsilon =  \int \varepsilon \, \Theta  (\varepsilon, \varepsilon_F) \tilde {n}_i (\varepsilon) d \varepsilon \,,
\end{eqnarray}

Here we adopt a slightly different approach to relating the FOE and KPM formally by introducing
\begin{equation}
m_i^{FOE} (\varepsilon') =  \int f (\varepsilon, \mu) \delta (\varepsilon', \varepsilon) n_i (\varepsilon) d \varepsilon \,, \label{eq:mFOE}
\end{equation}
and
\begin{equation}
m_i^{KPM} (\varepsilon') =  \int \Theta (\varepsilon',\varepsilon_F) K (\varepsilon', \varepsilon) n_i (\varepsilon) d \varepsilon \,. \label{eq:mKPM}
\end{equation}

The energy and number of electrons may be obtained from
\begin{eqnarray}
U_{band,i}^{FOE} &=  \int \varepsilon \,m_i^{FOE} (\varepsilon) d \varepsilon \,,\,\,\,\, N_i^{FOE} =\int m_i^{FOE} (\varepsilon) d \varepsilon \,, \\�
U_{band,i}^{KPM} &=  \int \varepsilon \,m_i^{KPM} (\varepsilon) d \varepsilon \,,\,\,\,\, N_i^{KPM} =\int m_i^{KPM} (\varepsilon) d \varepsilon \,.
\end{eqnarray}
From Eq.(\ref{eq:mFOE}) and (\ref{eq:mKPM}) it is evident that the Fermi Operator Expansion and the kernel Polynomial Method converge to the same limit when $f (\varepsilon, \mu)$ becomes a step function at $T=0$K and the kernel $K (\varepsilon', \varepsilon)$  approaches the delta function as the number of moments in the expansion is increased.

Clearly at finite temperature or a finite number of moments FOE and KPM will be different; here we compare the two approaches to lowest order. We assume that we may identify $\mu= \varepsilon_F$ in expression (\ref{eq:mFOE}) and (\ref{eq:mKPM}), where $\varepsilon_F$ is the Fermi level in the KPM. Then, by taking the derivative w.r.t. $\mu$ and integrating over $\varepsilon'$ , the following identity has to hold
\begin{equation}
\frac {\partial f (\varepsilon, \mu)} {\partial \mu} =  K (\mu, \varepsilon) \,,
\end{equation}
if no further assumptions regarding the local density of states $n_i (\varepsilon)$ are made. As expected, we may understand the kernel as the width of the smearing function that is indicated by a non-zero first derivative. 

A systematic study of different smearing functions by Liang {\it et al.}\cite{Liang2003} has indicated that the most rapid convergence for a FOE is found by using a complementary error function as the electronic distribution function, i.e., 
\begin{equation}
f =\frac {1} {2} \left [ 1 - {\rm{erf}} \left ( - ( \varepsilon - \mu) / \gamma \right) \right ] \,,
\end{equation}
where the width $\gamma$ acts as an effective ``temperature''. In this case, 
\begin{equation}
\frac {\partial f} {\partial \mu} = \frac {1} {\sqrt {\pi\gamma^2}} \exp \left ( - \left ( \frac {\varepsilon - \mu} {\gamma} \right )^2 \right ) \,,\label{eq:gaussian_kernel}
\end{equation}
so that the underlying kernel is expected to be shaped like a Gaussian. The width of the Jackson Kernel, for example, is given by $\gamma = \sqrt{2} \pi/N_{\rm max}$ \cite{Weisse06} and thus $N_{\rm max}$ in KPM may be used to set the electron temperature in a simulation.

\section{Kernels and damping factors}
\begin{figure}
\centering
\includegraphics[width = 0.8\textwidth]{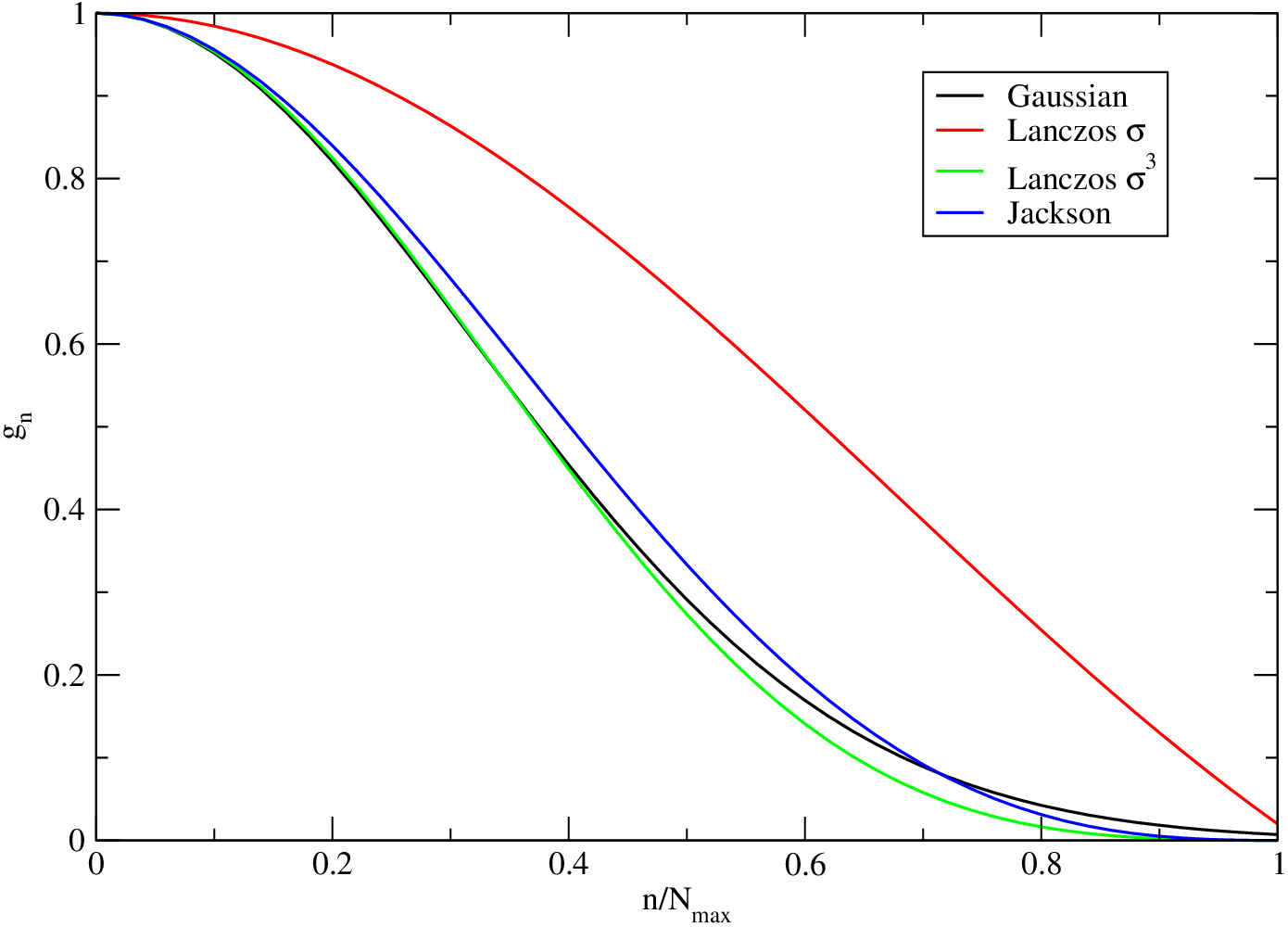}
\caption{Damping coefficients $g_n$ as a function of $n/N_{\rm max}$ for a number of kernels. The Jackson kernel\cite{jackson11} guarantees strictly positive densities of states, while the Lanczos $\sigma$ kernels\cite{lanczos1966} ($g_n^{(\sigma^M)} = \sinc ( \pi n/N_{max})^M$) were designed primarily to damp Gibbs oscillations in Fourier expansions.} \label{fig:damping_coefficients}
\end{figure}

In the following section we illustrate more closely the connection between the KPM and FOE, by comparing numerically the results obtained via the two approaches. We consider first an arbitrary function $f(x)$ (which could be the local density of states), for which we have evaluated the first $N_{\rm max}$ Chebyshev moments. A pure Chebyshev expansion of this function would be given by
\begin{equation}
f (x) \approx \frac{1}{\pi \sqrt {1 - x^2}} (\mu_0^T + 2 \sum_{n=1}^{N_{\rm max}} \mu_n^T T_n (x) ). \label{eq:cheb_expansion}
\end{equation}
Here $\{T_n (x) \}$ are the Chebyshev polynomials of the first kind, and the Chebyshev moments
\begin{equation}
\mu_n^T = \int_{-1}^{1} f(x) T_n (x) \diff x \label{eq:cheb_moments}
\end{equation}
At points where $f(x)$ has discontinuities, for example, in the region of a band gap, the expansion of Eq. (\ref{eq:cheb_expansion}) exhibits Gibbs oscillations, which may lead to regions where the density of states is negative. The use of a kernel in Eq. (\ref{eq:kpm_definition}) acts to ``damp'' higher-moment oscillations. This leads to the introduction of damping coefficients $\{ g_n\}$, so that Eq. (\ref{eq:cheb_expansion}) becomes
\begin{equation}
f (x) \approx \frac{1}{\pi \sqrt {1 - x^2}} (\mu_0^T + 2 \sum_{n=1}^{N_{\rm max}} g_n \mu_n^T T_n (x) ). \label{eq:cheb_damped_expansion}
\end{equation}
We now determine an approximate expression for the $\{ g_n \}$ for the case of the Gaussian kernel Eq. (\ref{eq:gaussian_kernel}).  In what follows, we assume that $\gamma \ll 1$, and $\mu \to 0$, as this will allow us to simplify a number of expressions. The real moments $\mu_{\rm Gauss}^{(n)}$ of this distribution are given by
\begin{equation}
\mu_{\rm Gauss}^{(n)} = \frac{1}{\sqrt{\pi}}\gamma^{n} \Gamma \left ( \frac{1 + n}{2} \right )
\end{equation}
for $n$ even. By assuming small $\gamma$, we can write the Chebyshev moments (of the first kind) $\tau_n$, by making use of the recurrence relations of the Chebyshev polynomials:
\begin{displaymath}
T_n (x) = 2 x T_{n-1} (x) - T_{n-2} (x)
\end{displaymath}
Thus, the first few (even) Chebyshev moments are given by:
\begin{eqnarray}
\tau_{\rm Gauss}^{(0)} & = & 1 \nonumber \\
\tau_{\rm Gauss}^{(2)} & = & \gamma^2 - 1 \nonumber \\
\tau_{\rm Gauss}^{(4)} & = & 6 \gamma^4 - 4 \gamma^2 + 1 \nonumber  \\
\tau_{\rm Gauss}^{(6)} & = & 60 \gamma^6 - 36 \gamma^4 + 9 \gamma^2 - 1 \nonumber \\
\tau_{\rm Gauss}^{(8)} & = & 840 \gamma^8 - 480 \gamma^6 + 120 \gamma^4 - 16 \gamma^2 +1 \label{eq:cheb_gauss}
\end{eqnarray}
If $\gamma \to 0$, $\delta_{\rm gauss}$ becomes simply the Dirac delta function, $\delta_{\rm Dirac}$, whose (even) Chebyshev moments are simply
\begin{eqnarray}
\tau_{\rm Dirac}^{(n)} = (-1)^{n/2} \label{eq:cheb_dirac}
\end{eqnarray}
For each $n$, the damping factor $g_n^{(\rm Gauss)}$ will be simply the ratio $\tau_{\rm Gauss}^{(n)}/\tau_{\rm Dirac}^{(n)}$. For small $n$, and since $\gamma \ll 1$, we can omit (approximately) terms of $\mathcal{O} (\gamma^4)$ and above, hence the damping factors become
\begin{equation}
g_n^{(\rm Gauss)} \approx 1 - \gamma^2 \left ( \frac{n}{2} \right )^2 \approx \exp \left ( - \left ( \frac{n\gamma}{2} \right )^2 \right ) \label{eq:damping_factor}
\end{equation}
These damping coefficients have been derived for a number of existing kernels\cite{Silver96}, and are illustrated in Fig. \ref{fig:damping_coefficients}. Of particular interest is the Jackson kernel, which guarantees a strictly positive density of states, while simultaneously minimising the broadening of the spectral function. 
In Fig. \ref{fig:damping_coefficients}, and throughout the rest of the paper, we choose the parameter $\gamma = \sqrt{2} \pi/N_{\rm max}$. This choice is motivated by Wei{\ss}e et al. \cite{Weisse06}, who have shown that the reproduction of the delta function via the Jackson kernel is a good approximation to a Gaussian of this width. It is noteworthy that, of the kernels considered here, the damping coefficients for the Gaussian kernel are the only ones which do not go to zero as $n \to N_{\rm max}$. 

\begin{figure}
\centering
\includegraphics[width = 0.8\textwidth]{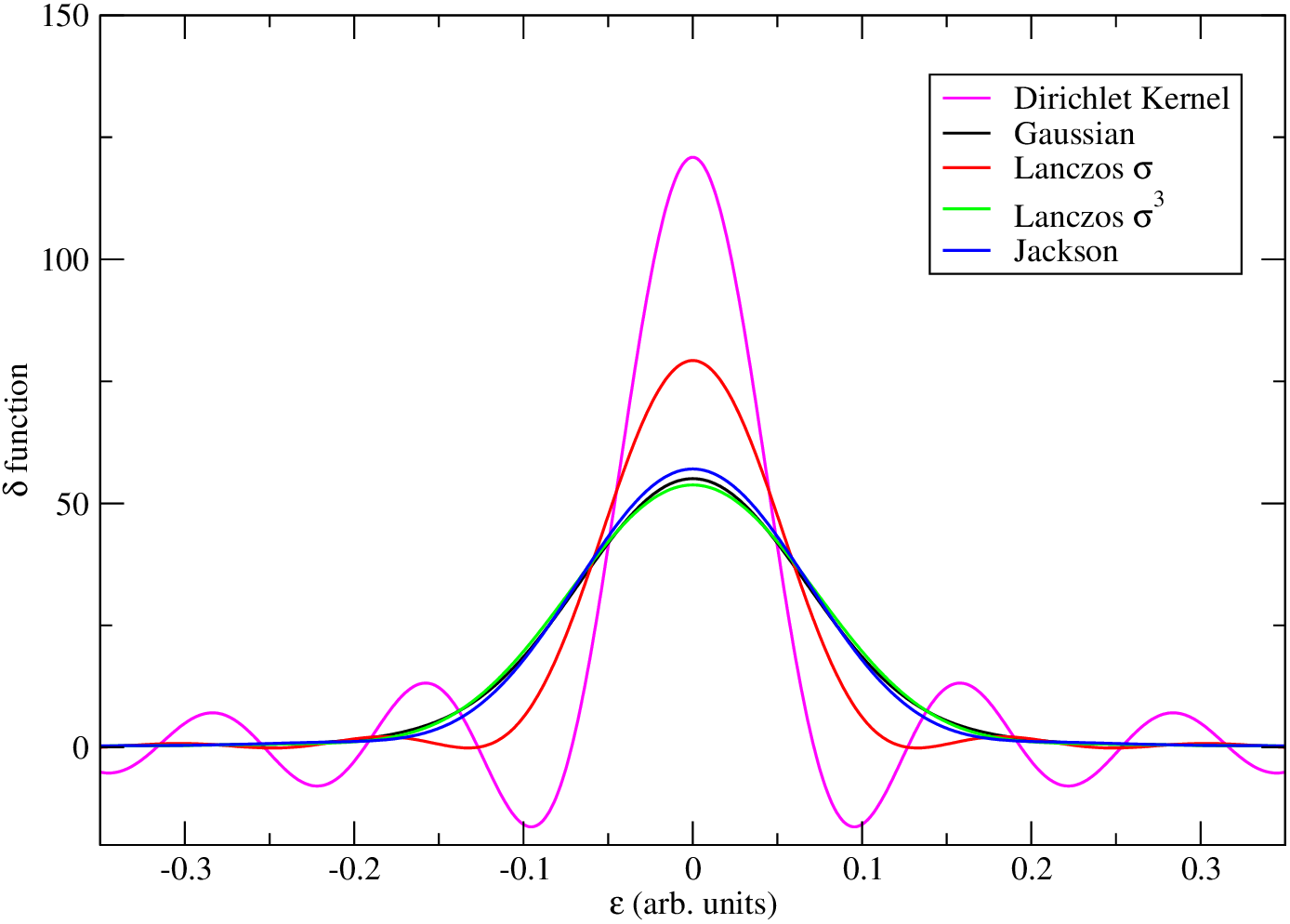} 
\caption{Expansion of the delta function for $N_{\rm max} = 50$ using the various kernels, as discussed in the text.}  \label{fig:delta_function}
\end{figure}

To demonstrate the effects of the smearing caused by the damping coefficients, the various kernels are applied to the problem of reproducing the delta function; the results are illustrated in Fig. \ref{fig:delta_function} for $N_{\rm max} = 50$. The Dirichlet kernel, which corresponds to the undamped expansion Eq. (\ref{eq:cheb_expansion}), demonstrates large Gibbs oscillations. The Jackson kernel, although significantly broadening the peak, has no negative values for the reproduced functions. The Lanczos $\sigma^3$ kernel gives similar results to the Jackson, without guaranteeing a strictly positive function. The Gaussian kernel, with $\gamma = \sqrt{2}\pi/N_{\rm max}$ is extremely similar to the Jackson kernel as anticipated. 

%

\section{Application to carbon}
Having established the connection between the KPM and FOE approaches, we move on to examine the performance of the two approaches when applied to real electronic structure models. As a test case, we have taken the orthogonal tight-binding (TB) model for carbon from Xu\cite{xu1992}, and examined the convergence of the band energy as a function of moments, for the various kernels. 

\begin{figure}
\centering
\includegraphics[width = 0.8\textwidth]{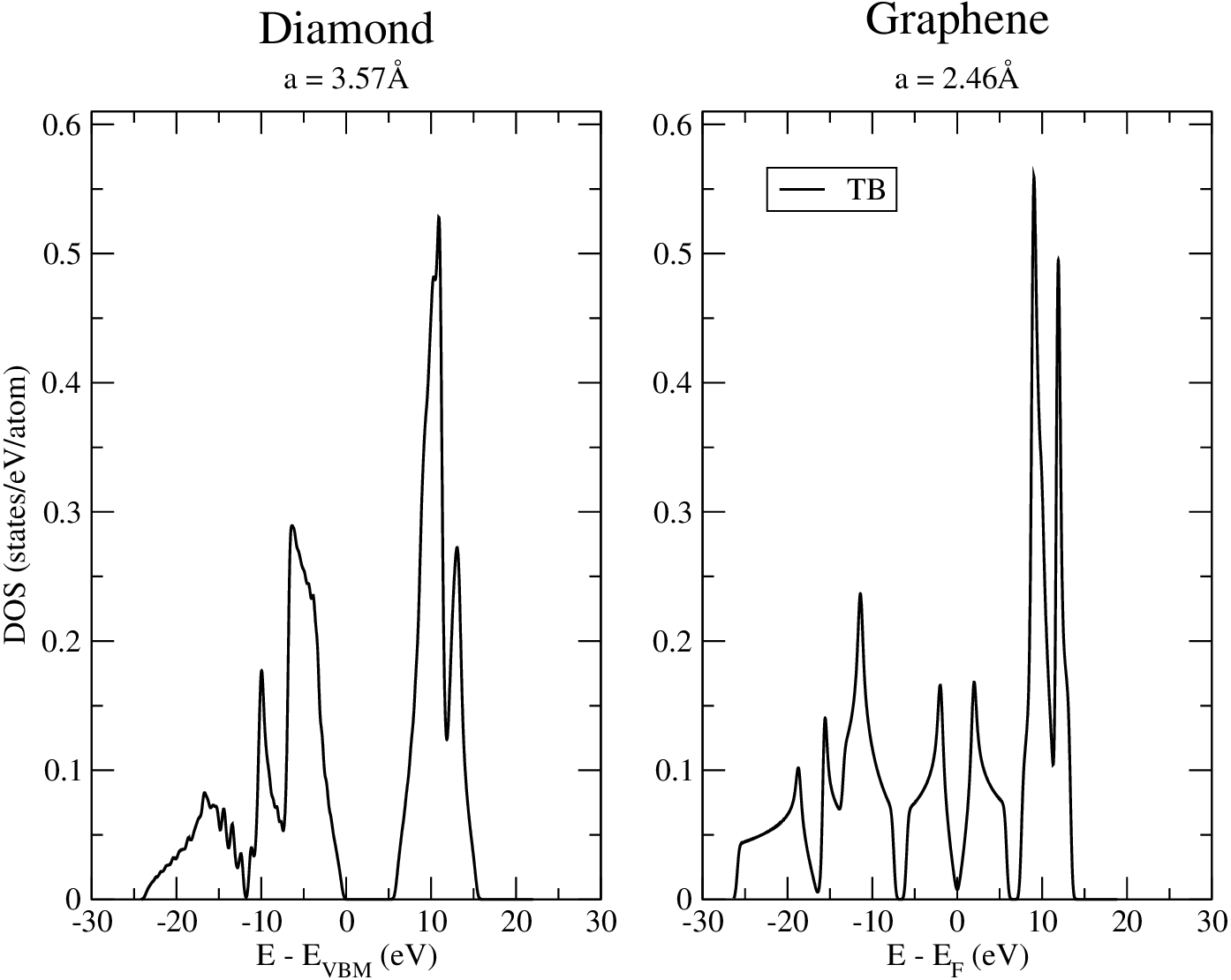}
\caption{Electronic densities of states for diamond and graphene at the equilibrium lattice constants, as evaluated within the tight-binding model of Xu. The zeroes of energy correspond to the valence band maximum ($E_{\rm VBM}$) for diamond, and the Fermi level ($E_{\rm F}$) for graphene.} \label{fig:carbon_tb_dos}
\end{figure}

The tight-binding electronic structures for carbon in the diamond and graphite structures  are illustrated in Fig \ref{fig:carbon_tb_dos}. The diamond structure exhibits a band-gap of $\sim 5$ eV, while the graphite structure has a semi-metallic DOS, with a narrow anti-resonant feature at the Fermi level. Due to the narrowness of these features in relation to the bandwidth of the system, we can anticipate that a relatively large number of moments are needed to accurately reproduce the total energy of the system.

\begin{figure}
\centering
\includegraphics[width = 0.7\textwidth]{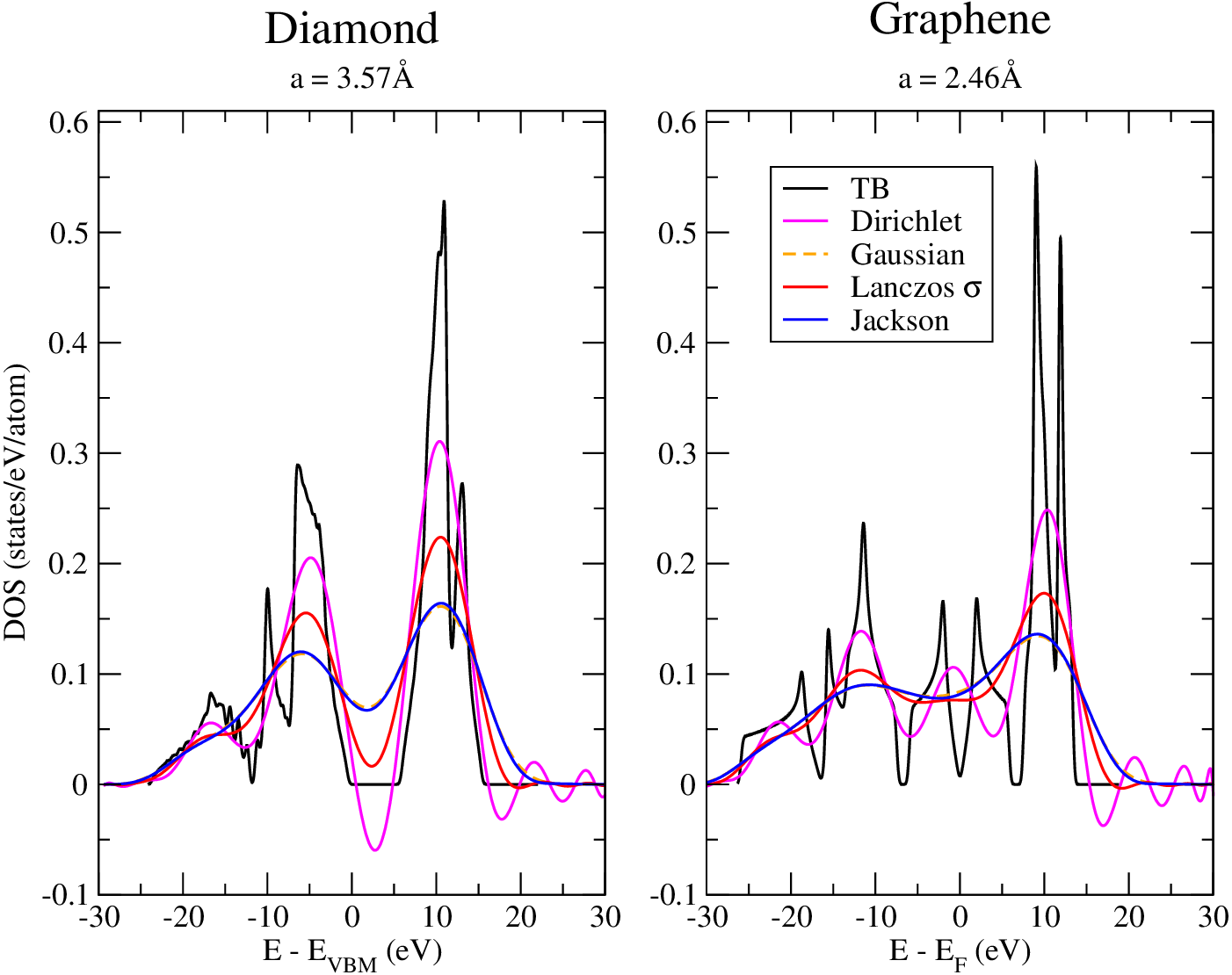}
\caption{Electronic densities of states for diamond and graphene evaluated for the various kernels, with $N_{\rm max} = 20$.} \label{fig:carbon_dos_20moments}
\end{figure}

\begin{figure}
\centering
\includegraphics[width = 0.7\textwidth]{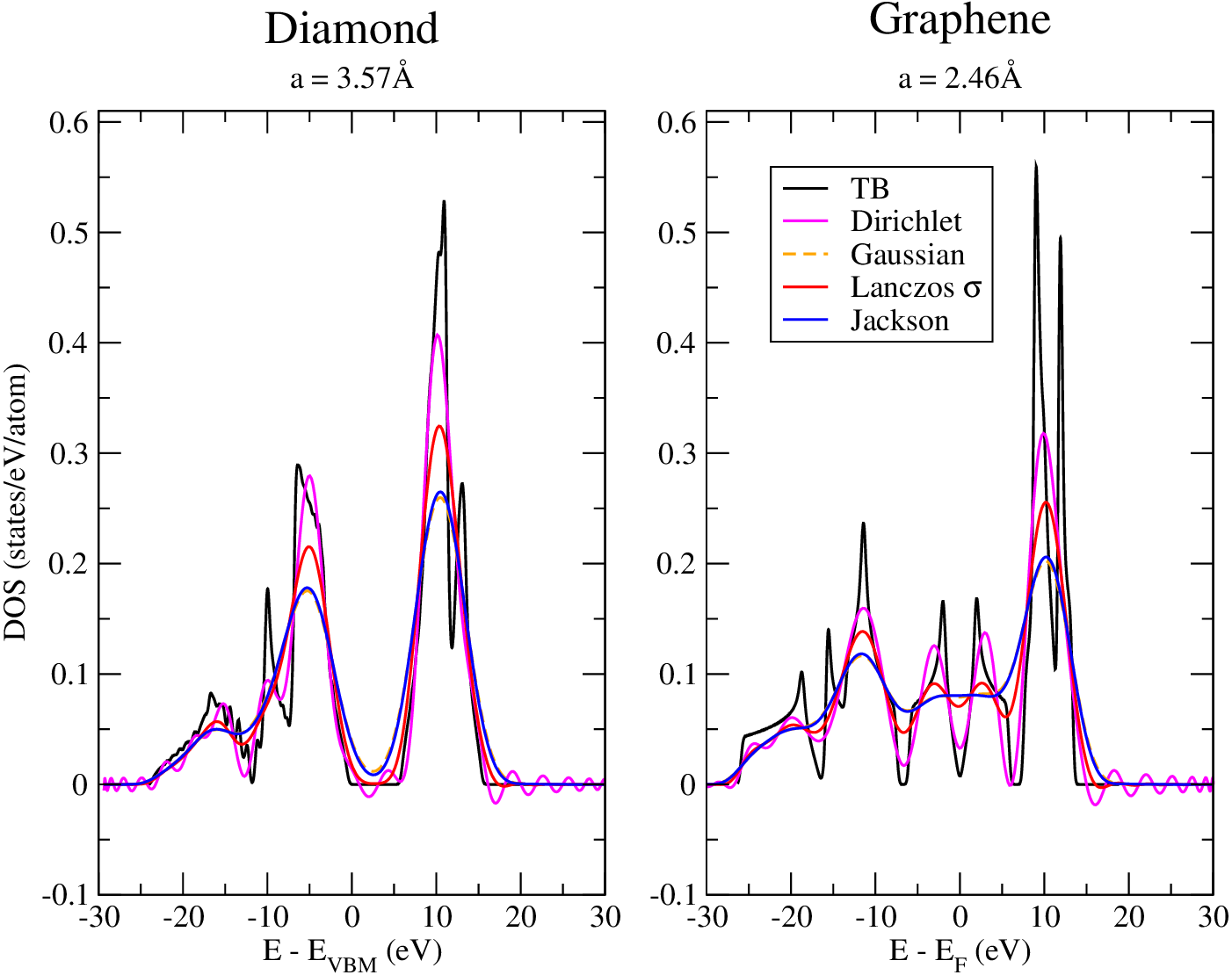}
\caption{Electronic densities of states for diamond and graphene evaluated for the various kernels, with $N_{\rm max} = 40$.} \label{fig:carbon_dos_40moments}
\end{figure}

In Figs. \ref{fig:carbon_dos_20moments} and \ref{fig:carbon_dos_40moments}, the densities of states of the diamond and graphite structures are shown for the various kernel approximations for $N_{\rm max} = 20$ and $N_{\rm max} = 40$ respectively. In Figure \ref{fig:carbon_dos_20moments}, the undamped Dirichlet kernel appears to show fast convergence, in particular in the reproduction of the band-gap of diamond. However, large Gibbs oscillations mean that the density of states for diamond has negative regions within the gap, which in turn results in the electron count being a multi-valued function of the energy and ambiguity in the determination of the valence band maximum. The Gaussian and Jackson kernels produce rather similar results, with differences only apparent toward the edge of the bandwidth. The Lanczos $\sigma$ kernel produces reasonable convergence, without producing negative DOS in the band-gap. In the case of graphene, it is clear that $N_{\rm max} = 20$ is insufficient to fully reproduce the complexity of the TB DOS, with only rather general features of the electronic structures being recovered. In Figure \ref{fig:carbon_dos_40moments}, the increased number of moments provides much more realistic reproduction of the TB DOS. In the case of diamond, all kernels produce a  reasonable description of the electronic structure; however, the undamped case exhibits significant Gibbs oscillations in the gap region. In the case of graphene, the broadening introduced by the Gaussian and Jackson kernels removes much of the local structure around the Fermi level, while the Lanczos $\sigma$ kernel reproduces much of the local structure with the only negative regions of the DOS appearing at the band edges. 

\begin{figure}
\centering
\includegraphics[width = 0.45\textwidth]{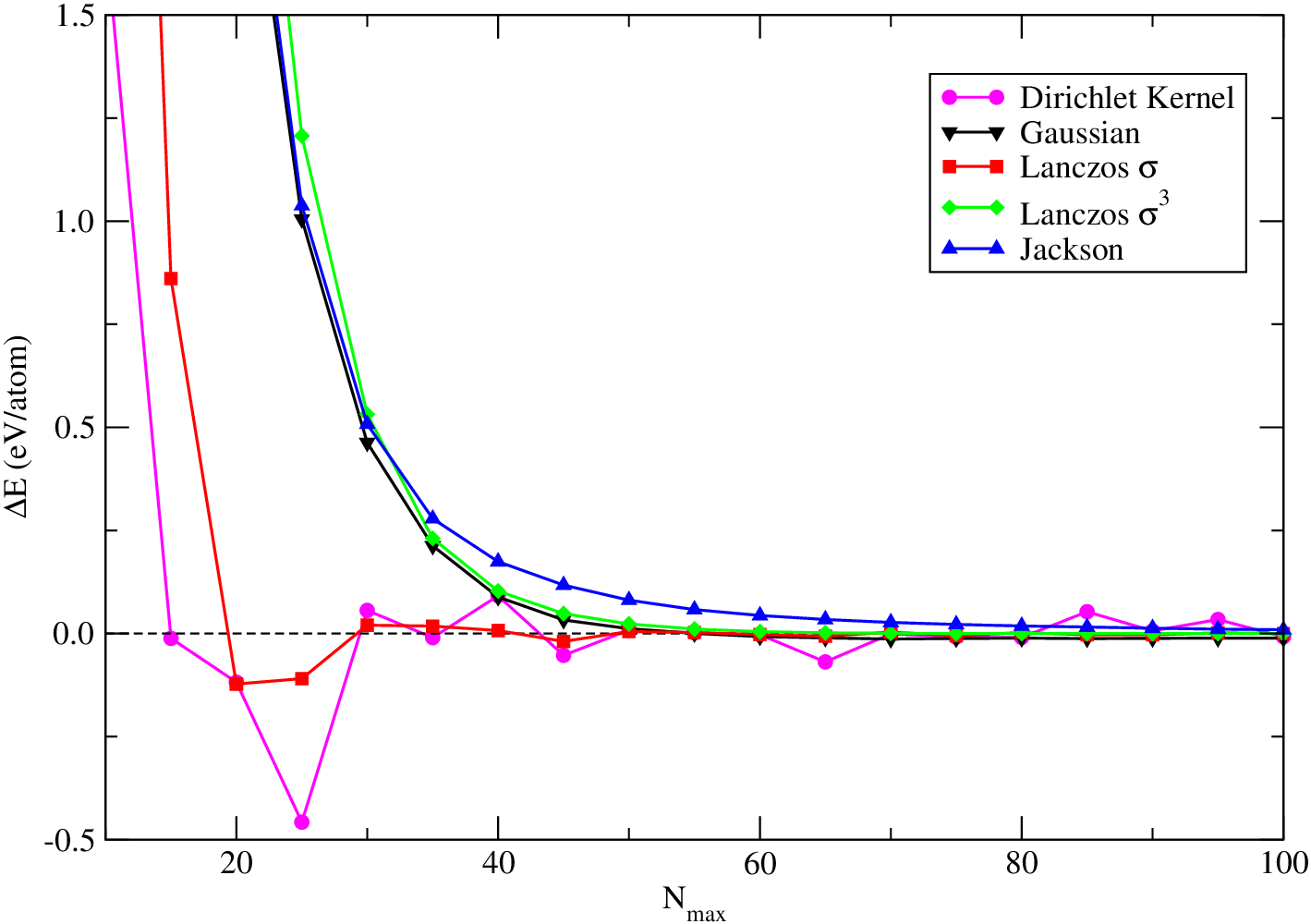}
\includegraphics[width = 0.45\textwidth]{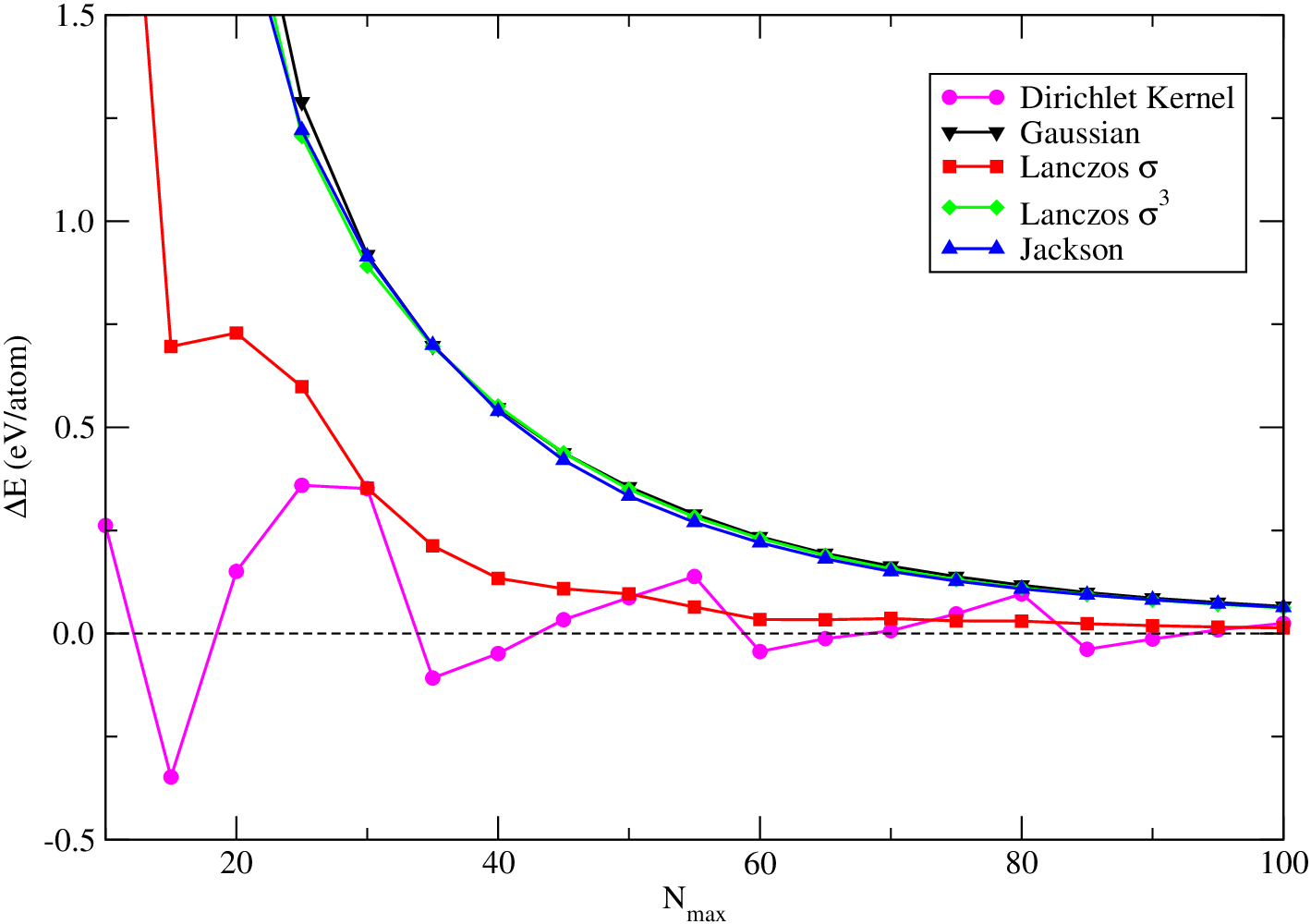}
\caption{Error $\Delta E$ in the band energy of the diamond (left) and graphene (right) as a function of $N_{\rm max}$ for the various kernels.} \label{fig:carbon}
\end{figure}

In Figure \ref{fig:carbon}, the error in the band energy (compared to exact diagonalisation) is shown as a function of the number of moments, for both diamond and graphene. The Dirichlet kernel, although closest at low moments, oscillates significantly around the true energy in both cases. The Gaussian, Lanczos $\sigma^3$ and Jackson kernels all converge monotonically from above to the exact value. As expected, the convergence for graphite is slower with $N_{\rm max}$, with $N_{\rm max} \approx 100$ required to have energy convergence $\sim 50$ meV/atom for the Gaussian, Lanczos $\sigma^3$ and Jackson kernels.

\begin{figure}
\centering
\includegraphics[width = 0.7\textwidth]{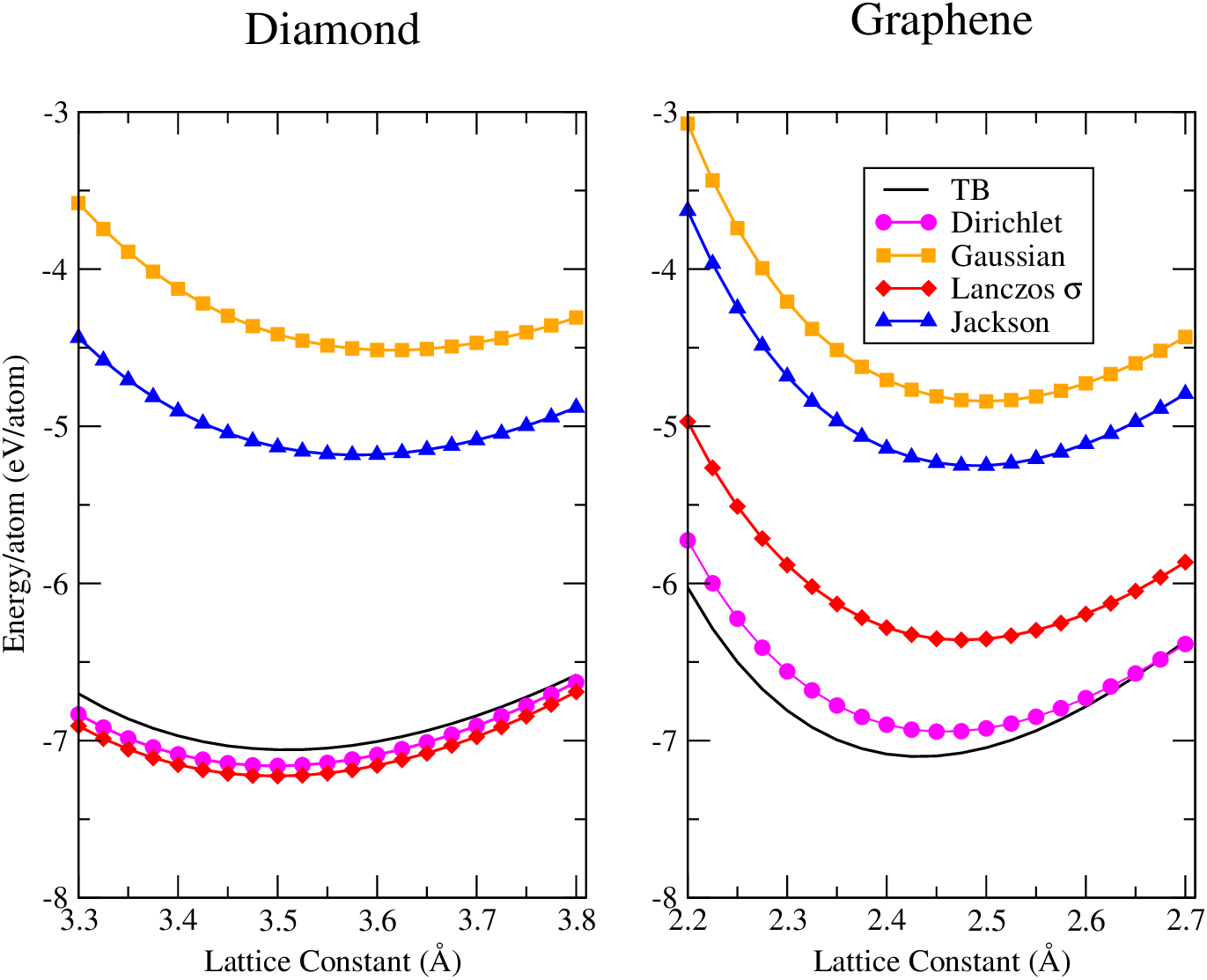}
\caption{Energetics of diamond and graphene as a function of lattice parameter as evaluated for the various kernels, with $N_{\rm max} = 20$.} \label{fig:energies_20moments}
\end{figure}

\begin{figure}
\centering
\includegraphics[width = 0.7\textwidth]{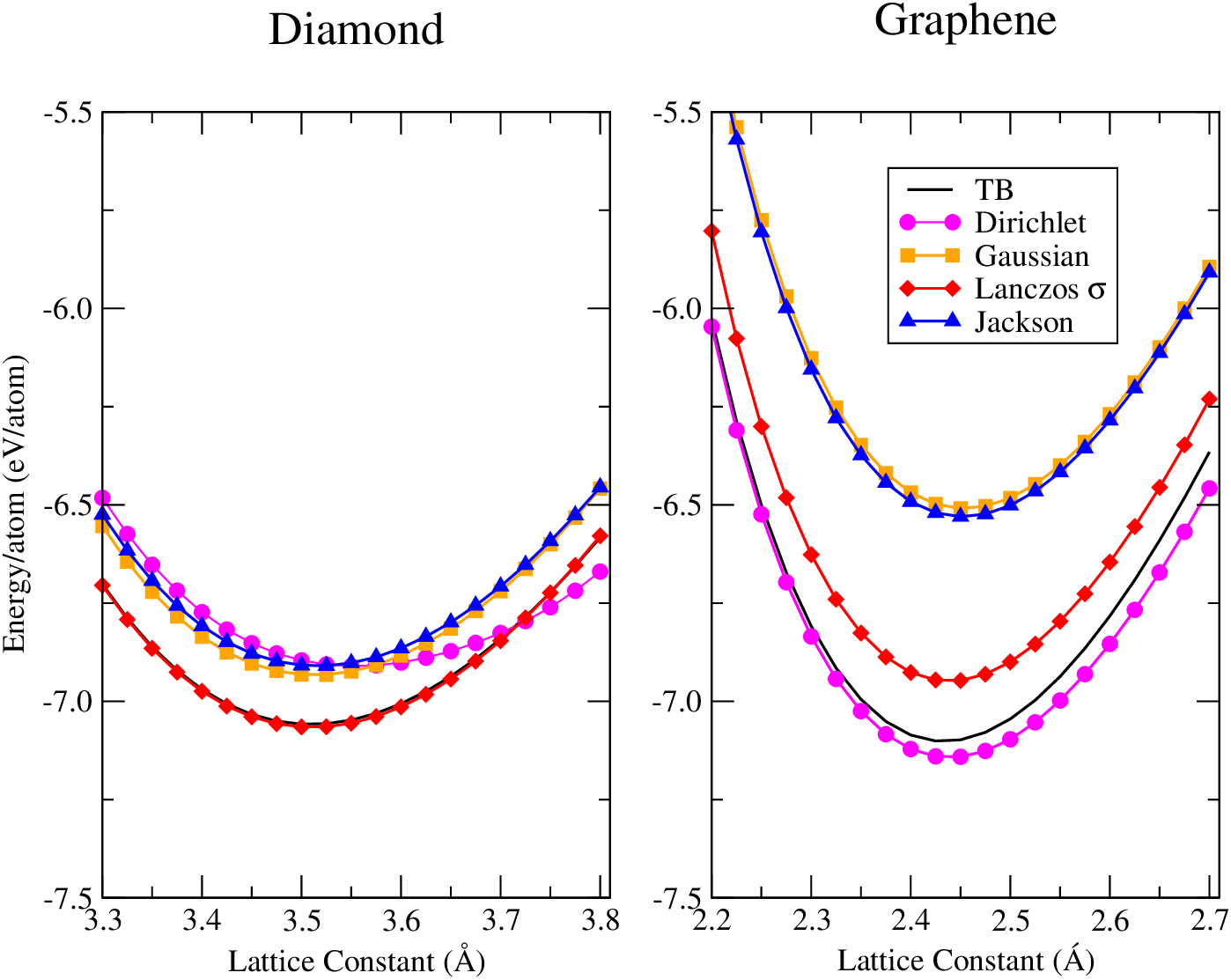}
\caption{Energetics of diamond and graphene as a function of lattice parameter as evaluated for the various kernels, with $N_{\rm max} = 40$.} \label{fig:energies_40moments}
\end{figure}

The non-variational behaviour of the Dirichlet and Lanczos $\sigma$ approaches becomes more clear when full energy-volume curves are evaluated. In Figure \ref{fig:energies_20moments}, the energies of diamond and graphene as a function of lattice constant are shown for the various approaches (with $N_{\rm max} = 20$) and compared with well-converged TB results. In the case of diamond, the Dirichlet and Lanczos $\sigma$ kernels undershoot the TB energies, with a reasonable prediction of the lattice constant and bulk modulus. The much smoother Gaussian and Jackson kernels give significantly higher energies, ostensibly due to an elevated effective electronic temperature. In the case of graphene, both the Dirichlet and Lanczos $\sigma$ kernels have energies above the TB result, thus falsely stabilising the diamond structure over the graphene. Even though the energy error for the Jackson kernel is of the order of $2$ eV, the energetic ordering between the two crystal structures is in good agreement with the TB result. In Fig \ref{fig:energies_40moments}, the results for $N_{\rm max} = 40$ are shown. In the case of diamond, all kernels give a reasonable description of the lattice constant and bulk modulus, with the exception of the Dirichlet kernel, which arises from the multivaluedness of the electron number due to the negative DOS within the band gap. For graphene, the slower convergence in energy with $N_{\rm max}$ is apparent even at $N_{\rm max} = 40$, with all damped kernels stabilising the diamond structure over graphene. In order to reliably determine energy differences, it is essential that the key features of the density of states are reproduced; the smearing out of the anti-resonance feature at the Fermi level in graphene has a significant effect on the total energy, even if the lattice constant is in fair agreement with the TB reference.

The results here show the ability of the KPM to reproduce, in a smooth and rigorous manner, the total energy of these systems, as well as the challenges of such an approach to reproduce narrow spectral features which drive small energy differences between structures. In a realistic calculation, the number of moments required to reliably describe the allotropes of carbon would be prohibitively expensive to evaluate exactly, thereby necessitating a means of estimating higher Chebyshev moments from lower ones. One recent approach\cite{Seiser13} is to estimate these from the Krylov subspace generated by Lanczos recursion approaches to the same problem. 

\section{Conclusion}
A connection between the KPM and the FOE approach to linear-scaling electronic structure calculations has been made. We have shown that the width of the Kernel in the KPM may be related directly to the electronic temperature. The use of a complementary error function as an electronic smearing function in the FOE is consistent with assuming a Gaussian Kernel within the KPM. Moreover, while in the FOE, the expansion coefficients $\{ c_k \}$ of Eq. (\ref{eq:poly}) must be re-evaluated for each change in chemical potential, this is not necessary within the KPM. Therefore, while the direct FOE and Gaussian kernel KPM method give virtually identical answers, the KPM approach is computationally more efficient.

\section*{References}

\end{document}